\documentclass[12pt,a4paper,reqno]{article}
\usepackage{graphics}
\usepackage{epsfig}

\usepackage{amsmath,amsthm,amssymb}
\newtheorem{theorem}{Theorem}[section]

\newtheorem{example}[theorem]{Example}
\theoremstyle{definition}

\theoremstyle{remark}

\numberwithin{equation}{section}

\begin{document}
\title{  Codes from the Incidence Matrices  of a zero-divisor Graphs}
%\thanks{The first author would like to thank the Department of Science and Technology (DST), New Delhi, India for their financial support
%in the form of INSPIRE Fellowship
%(DST Award Letter No.IF130493/DST/INSPIRE Fellowship / 2013/(362) Dated: 26.07.2013) 
%to carry out this work.}

\author{N. Annamalai\\
Assistant Professor\\
%\thanks{Corresponding Author}\\
Indian Institute of Information Technology Kottayam\\
Pala-686635, Kerala, India\\
{Email: algebra.annamalai@gmail.com}
\bigskip\\
C Durairajan\\
Associate  Professor\\
Department of Mathematics\\ 
School of Mathematical Sciences\\
Bharathidasan University\\
Tiruchirappalli-620024, Tamil Nadu, India\\
{Email: cdurai66@rediffmail.com}
%Patrick Sol\'e \\
%CNRS, I3S \\
%ESSI, BP 145 \\
%Route des Colles\\
%06 903 Sophia Antipolis \\
%France \\
%{Email: sole@essi.fr}
%}
\hfill \\
\hfill \\
\hfill \\
\hfill \\
{\bf Running head:} Codes from the Incidence Matrices  of a zero-divisor Graphs}
\date{}
\maketitle

\newpage

\vspace*{0.5cm}
%\vspace*{1cm}
\begin{abstract} 
In this paper,	we examine the linear codes with respect to the Hamming metric from incidence matrices of the zero-divisor graphs with vertex set is the set of all non-zero zero-divisors of the ring $\mathbb{Z}_n$ and two distinct
vertices being adjacent iff their product is zero over $\mathbb{Z}_n.$ The main parameters of the codes are obtained.
\end{abstract}
%\vspace{1.5cm}
\vspace*{0.5cm}
%%%%%%%%%%%%%%%%%%%%%%%%%%%%%%%%%%%%%%%%%


\vspace{0.5cm}
\vspace{1.5cm}
%\newpage
\section{Introduction}
The study of zero-divisor graphs has received a lot of attention in recent years.
The concept of associating zero-divisors of a ring $R$ with a graph was first introduced by Beck\cite{beck}. A simple graph $G$ with vertex set as the set of all zero-divisors of the ring $R$ is said to be a {\it zero-divisor graph } if two vertices $x$ and $y$ are adjacent iff $x\cdot y = 0$ in $R.$ 
%The graph $G$ is known as the zero-divisor graph of the ring $R.$
Anderson and Livingston\cite{and} were the first who
simplify Beck's zero-divisor graph. Their motivation was to give a better clarification of the zero-divisor structure of the ring. The new zero-divisor graph is still a simple graph, only the non-zero zero-divisors of the commutative ring are
considered. They defined the zero-divisor graph $\Gamma(R)$ as a graph with vertex
set $Z^*(R),$ i.e., the set of non-zero zero-divisors of the ring $R$ such that
two distinct vertices $x$ and $y$ are adjacent if and only if $xy = 0.$ In 2002, Redmond\cite{red} extended the definition of zero-divisor graph to non-commutative
rings.

Codes from the row span of incidence matrix or adjacency matrix of 
various  graphs are discussed in \cite{dan, cd1, cd2}.

\section{Preliminaries }
\label{intro}
\indent Let $\mathbb{F}_q$ denote the finite field with cardinality $q.$  Then the Hamming weight 
$w_{H}(a$) of $a\in \mathbb{F}_q$ is defined by \begin{align*}
	w_{H}(a)=\begin{cases}0&\,\,\text{if}\,\,a=0\\1&\,\,\text{if}\,\, 
		a\neq 0
	\end{cases}
\end{align*}
For any two elements  $a,b\in\mathbb{F}_q$,
the Hamming distance $d_H(a,b)=w_{H}(a-b).$ Note that  $d_H(a, b)=d_H(b, a).$
%Let $\mathbb{F}_q^n$ be a vector space over $\mathbb{F}_q$ of dimension $n.$ 
Let $x=(x_1, \dots, x_n)\in \mathbb{F}_q^n,$ then the Hamming weight $w_{H}(x)$ of $x$ is defined by
$$w_{H}(x)=\sum\limits_{i=1}^nw_{H}(x_{i}).$$
For any 
$x=(x_1,\dots, x_n)$ and $y=(y_1, \dots, y_n)$ in $\mathbb{F}_q^n,$ the 
Hamming distance between $x$ and $y$ is defined by 
$$d_H(x,y)=\sum\limits_{i=1}^{n}d_H(x_i,y_i)=\sum\limits_{i=1}
^{n}w_H(x_i-y_i).$$
\indent A non-empty subset $C$ of $\mathbb{F}_{q}^{n}$ is said to be a $q$-ary 
code of length $n.$ An element of the code
$C$ is called a codeword. A $q$-ary linear code $C$ of length $n$  is a 
subspace of the 
vector space $\mathbb{F}_{q}^{n}$ over 
$\mathbb{F}_{q}.$ The minimum Hamming distance of a code $C$ is defined by
$$
d_{H}(C)=\min\limits_{c_{1},c_{2}\in C}\{
d_{H}(c_{1}, 
c_{2}) \mid c_1\neq c_2\}.$$
The minimum weight of a code $C$ is the smallest among all weights of the non-zero codewords of $C.$ For $q$-ary linear code, we have 
$d_{H}(C)=w_{H}(C).$ For basic coding theory, we refer \cite{san}.

All the codes here are linear codes and the notation $[n, k, d]_{q}$ 
will be used for a $q$-ary linear code of length $n,$ dimension $k$ and minimum 
distance $d.$ A generator matrix $G$ for a 
linear code $C$ is a matrix whose rows form a basis for the subspace $C$ and $C_{q}(G)$ is a code generated by matrix $G$ over a finite field $\mathbb{F}_{q}$ and dimension of the code $C_{q}(G)$ is the rank of the matrix $G$ over $\mathbb{F}_{q}.$\\

Let $\Gamma=(V, E)$  be a graph with vertex set $V,$ edge set $E$ and for any $x, y \in V,\,[x,y]$ is denoted by an edge between $x$ and $y.$ An incidence matrix of 
$\Gamma$ is a $|V|\times|E|$ matrix $B$ with rows labelled by the vertices and 
columns by the edges and entries $b_{ij}=1$ if the vertex labelled by row $i$ 
is incident with the edge labelled by column $j$ and $b_{ij}=0$ otherwise.

Let $W, X\subseteq V$ with $W\cap X=\emptyset$ and let  $E(W, X)$ be the set of edges that have one end in $W$ and the other end in $X.$ Write $|E(W, X)|=q(W, X).$ 

An edge-cut of a connected graph $\Gamma$ is the set $S\subseteq E$ such that $\Gamma- S=(V, E-S)$ is disconnected.

The edge-connectivity $\lambda(\Gamma)$ is the minimum cardinality of an edge-cut. In fact, $$\lambda(\Gamma)=\min\limits_{\emptyset \neq W\subseteq V} q(W, V-W).$$
For any connected graph $\Gamma,$ we have $\lambda(\Gamma)\leq \delta(\Gamma)$
where $\delta(\Gamma)$ is minimum degree of the graph $\Gamma.$ 
\begin{theorem}\cite{reddy}\label{20}
	The edge connectivity of $\Gamma(\mathbb{Z}_n)$ for $n = p_1^{\alpha_1}p_2^{\alpha_2}\cdots p_k^{\alpha_k}$ is $r = \min\limits_{ 1 \leq i \leq k}\{p_i -1 \}.$
\end{theorem}

We denote the code generated by the rows of the incidence matrix $G$ of the  graph $\Gamma$ is by $C_p(G)$ over the finite field $\mathbb{F}_p.$
\begin{theorem}\cite{dan}\label{21}
	\begin{itemize}
		\item[1.] Let $\Gamma = (V, E)$ be a connected graph and let  $G$ be a $|V|\times|E|$ incidence matrix for $\Gamma.$ Then the main parameters of the code	$C_2(G)$ is $[|E|, |V|- 1, \lambda(\Gamma)]_2.$
		\item[2.]  Let $\Gamma = (V, E)$ be a connected bipartite graph and let $G$ be a $|V|\times|E|$ incidence matrix for $\Gamma.$ Then the incidence matrix generates 
		$[|E|, |V|-1,\lambda(\Gamma)]_p$ code for odd prime $p.$
	\end{itemize}
\end{theorem}
In this paper, we study a code from the incidence matrix of the zero-divisor graph 
$\Gamma(\mathbb{Z}_n)$ over $\mathbb{F}_p$ and we determined the parameters of the code. In Section 3, we discussed a code from incidence matrix of a zero-divisor graph $\Gamma(\mathbb{Z}_{p_1p_2})$ over the field $\mathbb{F}_p.$ In Section 4, we discussed a code from incidence matrix of a zero-divisor graph $\Gamma(\mathbb{Z}_{p_1p_2 p_3})$ over $\mathbb{F}_2.$ Finally, we studied the code from incidence matrix of the zero-divisor  $\Gamma(\mathbb{Z}_{p_1p_2 \cdots p_r})$ over  $\mathbb{F}_2$ in Section 5. All the zero-divisor graphs considered in this article are simple and undirected.

\section{ Code from Incidence Matrix of a Zero-divisor Graph $\Gamma(\mathbb{Z}_{p_1p_2})$}
In this section, we  study the codes obtained from the incidence matrix of the zero-divisor graph $\Gamma(\mathbb{Z}_{p_1p_2})$ over $\mathbb{F}_p$ and we find the parameters of the code.

Let $n=p_1p_2$ where $p_1$ and $p_2$ are distinct primes. Let $\mathbb{Z}_n$ be  the ring of integers modulo $n.$ Then the  zero-divisors of $\mathbb{Z}_n$  is $Z^*(\mathbb{Z}_n)=\{p_1, 2p_1, \cdots, (p_2-1)p_1, p_2, 2p_2,\cdots, (p_1-1)p_2\}.$

Let $\Gamma(\mathbb{Z}_n)$ be the graph with vertex set $V=Z^*(\mathbb{Z}_n)$ and two distinct vertices $a$ and $b$ are adjacent if and only if $ab = 0.$

Let $A=\{x p_1 \mid x=1, 2, \cdots, p_2-1\}$ and let $B=\{yp_2 \mid y=1, 2, \cdots, p_1-1\}$ be the two disjoint subsets of the vertex set $V.$ Then $|A|=p_2-1$ and $|B|=p_1-1.$ The zero-divisor graph $\Gamma(\mathbb{Z}_{n})$ can be viewed as follows: 

\begin{center}
	\includegraphics{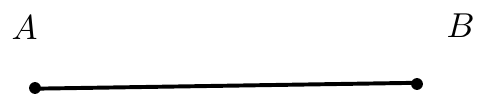}
\end{center}
That is, every vertex in $A$ is adjacent with all vertices in $B$ and vice versa. 
Clearly, $\Gamma(\mathbb{Z}_n)$ is a connected and complete bipartite graph with bipartition $A$ and $B$ and  $|A||B|=(p_1-1)(p_2-1)$ edges.
Then the incidence matrix of the zero-divisor graph $\Gamma(\mathbb{Z}_n)$ is

$$G=\begin{bmatrix}
	I_{p_2-1}&\vline&I_{p_2-1}&\vline&\cdots&\vline&I_{p_2-1}\\
	\hline
	{\bf 1}&\vline &{\bf 0}&\vline&\cdots&\vline&{\bf 0}\\
	{\bf 0}&\vline &{\bf 1}&\vline&\cdots&\vline&{\bf 0}\\
	\vdots&\vline&\vdots&\vline&\vdots&\vline&\vdots\\
	{\bf 0}&\vline &{\bf 0}&\vline&\cdots&\vline&{\bf 1}
\end{bmatrix}_{(p_1+p_2-2)\times ((p_1-1)(p_2-1))} $$
where $I_{p_2-1}$ is the $(p_2-1)\times (p_2-1)$ identity matrix, ${\bf 1}=[1 1 \cdots 1]_{1\times (p_2-1)}$ and ${\bf 0}=[0 0 \cdots 0]_{1\times (p_2-1)}.$

%In \cite{3}, they showed that the code generated by the incidence matrix of a connected graph is a $[|E|, |V|-1, \lambda(\Gamma)]_2$-code where $|E|$ is the length of the code, $|V|-1$ is the dimension of the code and $\lambda(\Gamma)$ is the minimum Hamming distance.
In \cite{reddy}, they proved the edge-connectivity $\lambda(\Gamma(\mathbb{Z}_{n}))$ of the zero-divisor graph $\Gamma(\mathbb{Z}_{n})$ is $\min \{(p_1-1),(p_2-1)\}.$ Therefore, by Theorem \ref{21}, the minimum distance of the code generated by the above incidence matrix $G$ is $\min \{(p_1-1),(p_2-1)\}$ and dimension $p_1+p_2-3.$ Thus, we have
\begin{theorem}
	Let $p_1$ and $p_2$ be two distinct primes. Then the main parameters of the $p$-ary linear code generated by the incidence matrix of the zero-divisor graph $\Gamma(\mathbb{Z}_{p_1p_2})$ is   $[(p_1-1)(p_2-1), p_1+p_2-3, \min\{p_1-1,p_2-1\}]$  for any prime $p.$
\end{theorem}
\begin{example}
	The zero-divisor graph $\Gamma(\mathbb{Z}_{15})$ is 
	\begin{center}
		\includegraphics{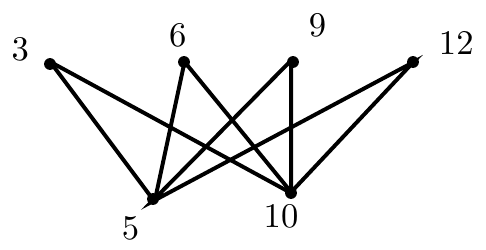}
	\end{center}
	Then the incidence matrix of the graph is 
	$$G=\begin{pmatrix}
		1&0&0&0&\vline&1&0&0&0\\
		0&1&0&0&\vline&0&1&0&0\\
		0&0&1&0&\vline&0&0&1&0\\
		0&0&0&1&\vline&0&0&0&1\\
		\hline
		1&1&1&1&\vline&0&0&0&0\\
		0&0&0&0&\vline&1&1&1&1\\
	\end{pmatrix}_{6\times 8}$$
	Since any five rows are linearly independent and the graph is bipartite, implies the dimension of the code $C_p(G)$ over the finite field $\mathbb{F}_p$ generated by $G$ is 5. The minimum distance of the code $C_p(G)$ is $2.$ Hence $C_p(G)$ is an $[8, 5, 2]_p$ code.
\end{example}

\section{ Code from Incidence Matrix of a zero-divisor Graph $\Gamma(\mathbb{Z}_{p_1p_2p_3})$}
In this section, we  study the code obtained from the incidence matrix of the zero-divisor graph $\Gamma(\mathbb{Z}_{p_1p_2p_3})$ over $\mathbb{F}_2$ and we find the main parameters of the code.

Let $n=p_1p_2p_3$ where $p_1, p_2$ and $p_3$ are distinct primes. 
Then the zero-divisors of $\mathbb{Z}_n$  can be partition into the following disjoints sets:
\begin{align*}
	A_1&=\{xp_1 \mid x=1, 2, \cdots, (p_2p_3-1), p_2\nmid x, p_3\nmid x\}\\
	A_2&=\{xp_2 \mid x=1, 2, \cdots, (p_1p_3-1), p_1\nmid x, p_3\nmid x\}\\
	A_3&=\{xp_3 \mid x=1, 2, \cdots, (p_1p_2-1), p_1\nmid x, p_2\nmid x\}\\
	A_4&=\{xp_1p_2 \mid x=1, 2, \cdots, (p_3-1)\}\\
	A_5&=\{xp_1p_3 \mid x=1, 2, \cdots, (p_2-1)\}\\
	A_6&=\{xp_2p_3 \mid x=1, 2, \cdots, (p_1-1)\}\\
\end{align*}
Then $|A_1|=(p_2-1)(p_3-1),$ $|A_2|=(p_1-1)(p_3-1),$ $|A_3|=(p_1-1)(p_2-1),$ $|A_4|=p_3-1,$ $|A_5|=p_2-1$ and $|A_6|=p_1-1.$ Hence $|V|=\sum\limits_{i=1}^{6} |A_i|.$
Then the zero-divisor graph $\Gamma(\mathbb{Z}_{n})$ can be viewed as follows:
\begin{center}
	\includegraphics{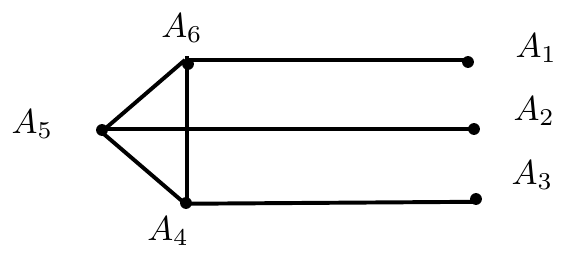}
\end{center}
where \includegraphics{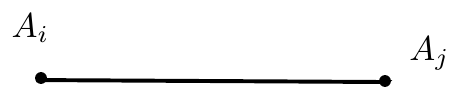} is a complete bipartite graph.
Then there are $|A_i||A_j|$ edges between the sets $A_i$ and $A_j.$ So the number of edges in the zero-divisor graph $\Gamma(\mathbb{Z}_{n})$ is $$|E|=|A_1||A_6|+|A_6||A_5|+|A_6||A_4|+|A_5||A_4|+|A_5||A_2|+|A_4||A_3|.$$
The set of edges between $A_i$ and $A_j$ is denoted by $[A_i, A_j].$
Clearly $\Gamma(\mathbb{Z}_n)$ is a connected graph. Then the incidence matrix of $\Gamma(\mathbb{Z}_n)$ is

$$G=\bordermatrix{&[A_1, A_6]&[A_2, A_5]&[A_3, A_4] &\vline&[A_4, A_5]&[A_4, A_6]&[A_5, A_6]&\cr
	A_1&{\bf 1} &  {\bf 0}  &{\bf 0}&\vline& {\bf 0} &{\bf 0}&{\bf 0}\cr
	A_2& {\bf 0}  &  {\bf 1} &{\bf 0}&\vline& {\bf 0} &{\bf 0}&{\bf 0}\cr
	A_3& {\bf 0} & {\bf 0} &{\bf 1}&\vline&{\bf 0} &{\bf 0}&{\bf 0}\cr
	\hline \cr
	A_4&{\bf 0} & {\bf 0} &{\bf 1}&\vline&{\bf 1} &{\bf 1}&{\bf 0}\cr
	A_5& {\bf 0} & {\bf 1} &{\bf 0}&\vline&{\bf 1} &{\bf 0}&{\bf 1}\cr 
	A_6& {\bf 1} & {\bf 0} &{\bf 0}&\vline&{\bf 0} &{\bf 1}&{\bf 1}}_{|V|\times |E|}$$
where the $(A_i, [A_k, A_j])$ entry is a $|A_i| \times |A_k| |A_j|$ all one matrix for $i = k \text{ or } i = j;$ zero matrix otherwise. Then the code generated by the incidence matrix $G$ is $C_2(G)$ with main parameters $[|E|, |V|-1, \lambda((\Gamma(\mathbb{Z}_{n})]_2.$
% where $|E|$ is the length of the code, $|V|-1$ is the dimension of the code and $\lambda((\Gamma(\mathbb{Z}_{n})$ is the minimum Hamming distance. 
Since the edge-connectivity $\lambda(\Gamma(\mathbb{Z}_{n}))$ of the zero-divisor graph $\Gamma(\mathbb{Z}_{n})$ is $\min \{(p_1-1),(p_2-1), (p_3-1)\}\cite{reddy},$ thus we have
\begin{theorem}
	Let $p_1, p_2$ and $p_3$ be two distinct primes. Then the code generated by the incidence matrix of the zero-divisor graph $\Gamma(\mathbb{Z}_{p_1p_2p_3})$ is a  $C_2(G)=[n, k, d]_2$  code over the finite field $\mathbb{F}_2$ where $n=|E|, k=|V|-1$ and $d=\min\{(p_1-1), (p_2-1), (p_3-1)\}.$
\end{theorem}
\begin{example}
	For $\mathbb{Z}_{30}$ with $p_1= 2, p_2 = 3$ and $p_3 = 5,$ we have
	\begin{align*}
		A_1 &= \{2, 4, 8, 14, 16, 22, 26, 28\}\\
		A_2 &= \{3, 9, 21, 27\}\\
		A_3 &= \{5, 25\}\\
		A_4 &= \{6, 12, 18, 24\}\\
		A_5 &= \{10, 20\}\\
		A_6 &= \{15\}.
	\end{align*}
	
	The zero-divisor graph $\Gamma(\mathbb{Z}_{30})$ is 
	\begin{center}
		\includegraphics{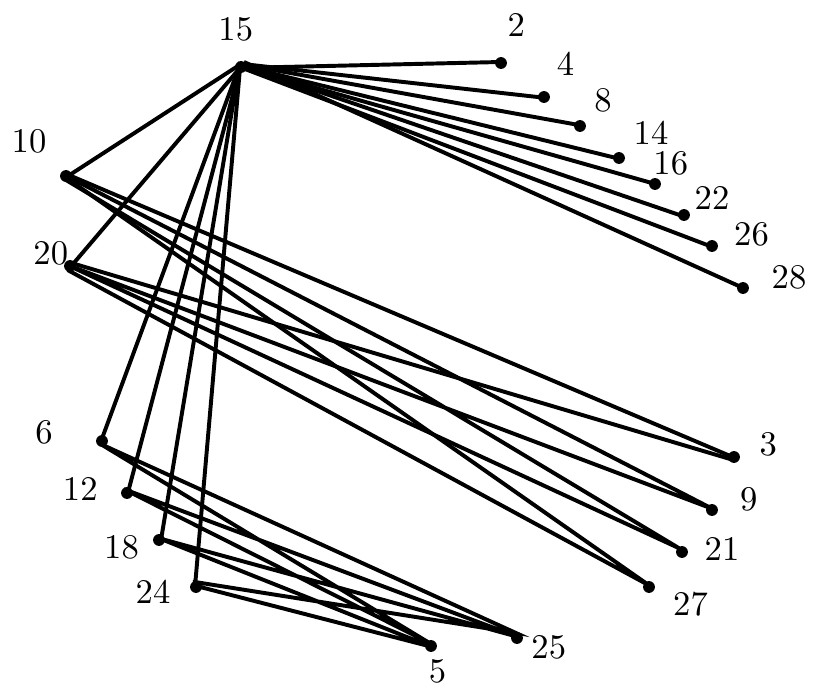}
	\end{center}
	Then the incidence matrix of the graph is 
	$$G=\bordermatrix{&[A_1, A_6]&[A_2, A_5]&[A_3, A_4] &\vline&[A_4, A_5]&[A_4, A_6]&[A_5, A_6]&\cr
		A_1&{\bf 1} &  {\bf 0}  &{\bf 0}&\vline& {\bf 0} &{\bf 0}&{\bf 0}\cr
		A_2& {\bf 0}  &  {\bf 1} &{\bf 0}&\vline& {\bf 0} &{\bf 0}&{\bf 0}\cr
		A_3& {\bf 0} & {\bf 0} &{\bf 1}&\vline&{\bf 0} &{\bf 0}&{\bf 0}\cr
		\hline \cr
		A_4&{\bf 0} & {\bf 0} &{\bf 1}&\vline&{\bf 1} &{\bf 1}&{\bf 0}\cr
		A_5& {\bf 0} & {\bf 1} &{\bf 0}&\vline&{\bf 1} &{\bf 0}&{\bf 1}\cr 
		A_6& {\bf 1} & {\bf 0} &{\bf 0}&\vline&{\bf 0} &{\bf 1}&{\bf 1}}_{21\times 38}$$
	where the $(A_i, [A_k, A_j])$ entry is a $|A_i| \times |A_k| |A_j|$ all one matrix for $i = k \text{ or } i = j;$ zero matrix otherwise.
	Since any five rows are linearly independent and the graph is bipartite, then the dimension of the code $C_2(G)$ over the finite field $\mathbb{F}_p$ generated by $G$ is 5. The minimum distance of the code $C_2(G)$ is $2.$ Hence $C_2(G)$ is an $[n, k, d]_2=[38, 20, 1]_2$ code.
\end{example}

\section{Codes from incidence matrix of a zero-divisor graph $\Gamma(\mathbb{Z}_{p_1p_2\cdots p_r})$}
In this section, we  study the codes obtained from the incidence matrix of the zero-divisor graph $\Gamma(\mathbb{Z}_{p_1p_2\cdots p_r})$ over $\mathbb{F}_2$ and we find the parameters of the code.

Let $n=p_1p_2\cdots p_r$ where $p_i's$ are distinct primes. Then the zero-divisors of $\mathbb{Z}_n$  can be partition into the following disjoints sets 
$$A_{p_i}=\{p_ix \mid x=1, 2, \cdots, (p_1p_2\cdots p_{i-1}p_{i+1}\cdots p_r)-1, p_j\nmid x , 1\leq j\leq r, i\neq j\}$$
$$A_{p_ip_j}=\{p_ip_jx \mid x=1, 2, \cdots, (p_1p_2\cdots p_{i-1}p_{i+1}\cdots p_{j-1})-1, p_s\nmid x , 1\leq s\leq r, i\neq s\neq j\}$$
In general,
$$A_{p_{1}p_{2}\cdots p_{i-1}p_{i+1}\cdots p_r}=\{x(p_{1}p_{2}\cdots p_{i-1}p_{i+1}\cdots p_r) \mid x=1,2,\cdots, p_i-1 \}.$$

%Then $|A_{p_i}|=\prod\limits_{{j=1},\\{i\neq j}}^{r}(p_j-1),$ $|A_{p_ip_j}|=\prod\limits_{{s=1},\\{s\neq i, j}}^{r}(p_s-1),$
If $G$ is the incidence matrix of the zero-divisor graph $\Gamma(\mathbb{Z}_{n}),$
then $C_2(G)$ is a $[|E|, |V|-1, \lambda((\Gamma(\mathbb{Z}_{n})]_2$ code where $|E|$ is the length of the code, $|V|-1$ is the dimension of the code and $\lambda((\Gamma(\mathbb{Z}_{n})$ is the minimum Hamming distance. Thus by Theorem \ref{20}, we have
%The edge-connectivity $\lambda(\Gamma(\mathbb{Z}_{n}))$ of the zero-divisor graph $\Gamma(\mathbb{Z}_{n})$ is $\min \{(p_1-1),(p_2-1), \cdots, (p_r-1)\}\cite{reddy}.$
\begin{theorem}
	Let $n=p_1p_2\cdots p_r$ where $p_i's$ are distinct primes. Then the code generated by the incidence matrix of the zero-divisor graph $\Gamma(\mathbb{Z}_{n})$ is a  $C_2(G)=[n, k, d]_2$  code over the finite field $\mathbb{F}_2$ where $n=|E|, k=|V|-1$ and $d=\min\{(p_1-1), (p_2-1), \cdots, (p_r-1)\}.$
\end{theorem}
\section*{Conclusion}
In this paper, we studied the codes generated by the incidence matrix of the zero-divisor graphs of the different commutative ring with unity. Also we found the main parameters of the code over different finite field. We have consider only simple and undirected graphs in this article. Finding the covering radius of these codes is the further direction to work.

\end{document}